\documentclass{article}
\usepackage[T1]{fontenc} 
\usepackage[utf8]{inputenc} 
\usepackage{ismir,amsmath,cite,url}
\usepackage{graphicx}
\usepackage{color}
\usepackage{lipsum}
\usepackage[bookmarks=false]{hyperref}
\usepackage{multirow, booktabs, bbding, enumerate, enumitem}%
\usepackage{subfigure}

\usepackage{lineno}

\title{SpecMaskGIT: Masked Generative Modeling of Audio Spectrograms for Efficient Audio Synthesis and Beyond}
\multauthor
{Marco Comunit{\`a}$^{*1,2}$ \hspace{0.2cm} Zhi Zhong$^{*2}$ \hspace{0.2cm}\thanks{*Equal contribution. Marco Comunit{\`a} was an intern at Sony.} Akira Takahashi$^2$ \hspace{0.2cm} Shiqi Yang$^2$ \hspace{0.2cm} Mengjie Zhao$^2$}
{
\bfseries{
Koichi Saito$^3$ \hspace{0.0cm} Yukara Ikemiya$^4$ \hspace{0.0cm} Takashi Shibuya$^4$ \hspace{0.0cm} Shusuke Takahashi$^2$ \hspace{0.0cm} Yuki Mitsufuji$^{2,3}$
}\\
$^1$ Queen Mary University of London, UK
$^2$ Sony Group Corporation, Japan\\
$^3$ Sony AI, US 
$^4$ Sony AI, Japan\\
}

\def\authorname{M. Comunit{\`a} and Z. Zhong}

\begin{document}

\maketitle

\begin{abstract}
\vspace{-1mm}
Recent advances in generative models that iteratively synthesize audio clips sparked great success to text-to-audio synthesis (TTA), 
but with the cost of slow synthesis speed and heavy computation. 
Although there have been attempts to accelerate the iterative procedure, high-quality TTA systems remain inefficient due to hundreds of iterations required in the inference phase and large amount of model parameters. 
To address the challenges, we propose SpecMaskGIT, a light-weighted, efficient yet effective TTA model based on the masked generative modeling of spectrograms. 
First, SpecMaskGIT synthesizes a realistic 10\,s audio clip by less than 16 iterations, an order-of-magnitude less than previous iterative TTA methods.
As a discrete model, SpecMaskGIT outperforms larger VQ-Diffusion and auto-regressive models in the TTA benchmark, while being real-time with only 4 CPU cores or even 30$\times$ faster with a GPU. 
Next, built upon a latent space of Mel-spectrogram, SpecMaskGIT has a wider range of applications (\textit{e.g.}, the zero-shot bandwidth extension) than similar methods built on the latent wave domain.
Moreover, we interpret SpecMaskGIT as a generative extension to previous discriminative audio masked Transformers, and shed light on its audio representation learning potential.
We hope our work inspires the exploration of masked audio modeling toward further diverse scenarios.
\end{abstract}
\vspace{-3mm}
\section{Introduction}
\label{sec:intro}
Text-to-audio synthesis (TTA) allows users to synthesize realistic audio and sound event signals by natural language prompts. 
TTA can assist the sound design and editing in the music, movie, and game industries, accelerating creators' workflow \cite{zhang2024musicmagus}. 
Therefore, TTA has earned arising attention in the research community. 

\begin{figure}[tb]
\centerline{\includegraphics[width=\linewidth]{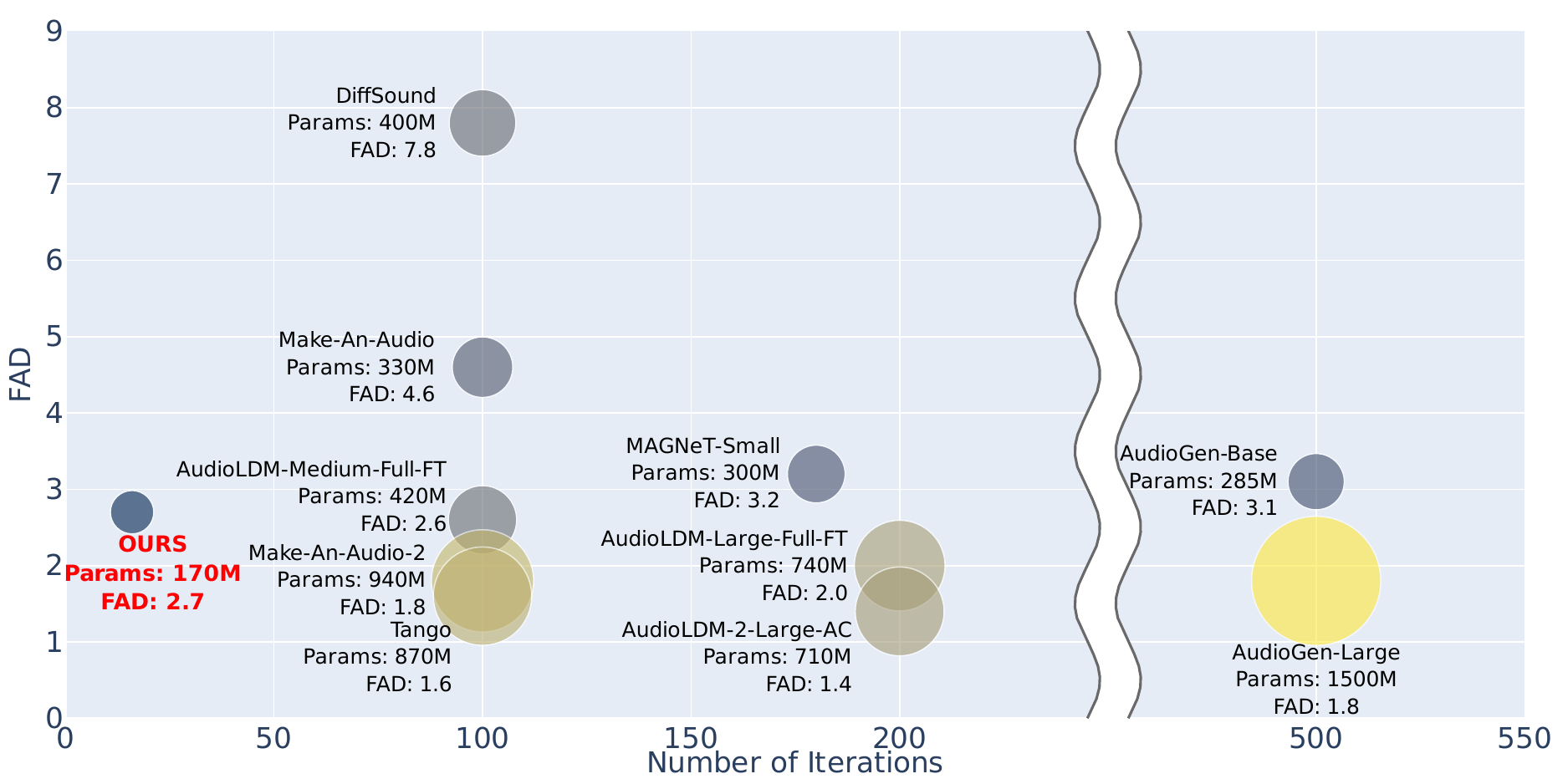}}
\vspace{-0.5cm}
\caption{Audio synthesis performance and number of synthesis iterations of different methods. The size of circle represents the model size. SpecMaskGIT achieves decent quality with only 16 iterations and a small model size.}
\label{fig:fad_numiter_size}
\vspace{-0.3cm}
\end{figure}
\begin{figure}[tb]
\centerline{\includegraphics[width=0.95\linewidth]{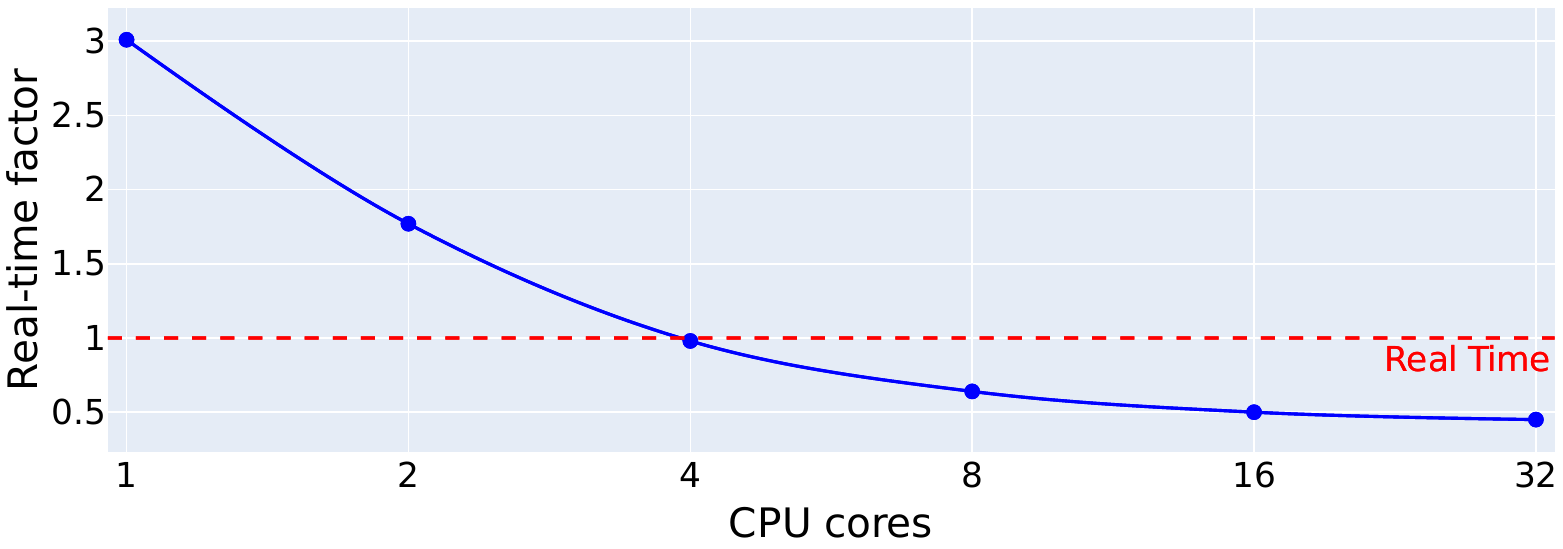}}
\vspace{-0.5cm}
\caption{Real-time factor of SpecMaskGIT on different Xeon CPU cores with standard Python implementation.}
\label{fig:runtime_cpu}
\vspace{-0.4cm}
\end{figure}

Recent advances in deep generative models, especially iterative methods such as diffusion \cite{yang2023diffsound, huang2023makeanaudio, liu2023audioldm, comunita2024syncfusion} and auto-regressive models \cite{kreuk2022audiogen, borsos2023audiolm, Li2023SJTU_AR}, have brought significant success to the sound quality and controllability in TTA tasks,
but with the cost of slow synthesis speed. 
Since the synthesis speed of iterative methods is dominated by the number of iterations required at inference,
techniques have been introduced to to reduce iterations, \textit{e.g.}, higher compression rate of raw audio signals\cite{kreuk2022audiogen} or more efficient diffusion samplers \cite{liu2023audioldm, huang2023makeanaudio2}. 
Nevertheless, these iterative methods remain slow in synthesis speed and demanding for computing resources, as they typically require hundreds of iterations to synthesize a short audio clip. Moreover, the runtime of a single iteration increases due to the huge model size.

To further improve the efficiency of audio synthesis, Garcia \textit{et al.} introduced the MaskGIT \cite{chang2022maskgit} synthesis strategy from computer vision to the realm of audio and proposed VampNet \cite{garcia2023vampnet}. 
Although VampNet can inpaint a 10-second clip with 24 iterations, 6 seconds are needed on GPU \cite{garcia2023vampnet}, which is still heavy for non-GPU environments. 
Moreover, VampNet is not compatible with text prompts or TTA tasks. 
Concurrent to our work, MAGNeT extended VampNet to text-conditional audio synthesis \cite{ziv2024magnet}. 
However, the method is less efficient as it requires 180 iterations, which is even heavier than some diffusion models that only requires 100 iterations \cite{liu2023audioldm, ghosal2023tango, liu2023audioldm2, huang2023makeanaudio2}. 
Since both VampNet and MAGNeT work in a wave-domain latent space, it is difficult to conduct frequency-domain inpainting tasks such as bandwidth extension (BWE) in a zero-shot manner. 
Besides the aforementioned limitations, the audio representation learning potential of a masked generative Transformer has not been investigated yet.

As a summary, an audio synthesis method that is compatible with text prompts, highly efficient in synthesis speed, and flexible for various downstream tasks is yet to be explored. 
To this end, we propose SpecMaskGIT, an efficient and flexible TTA model based on the masked generative modeling of audio spectrograms, to address the above challenges.
Our contributions lie in the following aspects.
\begin{itemize}[leftmargin=*]
\vspace{-2mm}
\item \textbf{Efficient and effective TTA.} SpecMaskGIT synthesizes a realistic 10-second audio clip by less than 16 iterations, which is one order-of-magnitude smaller than previous iterative methods shown in Fig.~\ref{fig:fad_numiter_size}. 
As a discrete generative model, SpecMaskGIT outperforms larger VQ-Diffusion (DiffSound \cite{yang2023diffsound}) and auto-regressive  (AudioGen-base \cite{kreuk2022audiogen}) models in a TTA benchmark, while being real-time with 4 CPU cores shown in Fig.~\ref{fig:runtime_cpu} or even 30$\times$ faster on a GPU. 
\vspace{-2mm}
\item \textbf{Flexibility in downstream tasks.} 
SpecMaskGIT is interpreted and implemented as a generative extension to previous discriminative audio masked Transformers \cite{koutini2021patchout, ntt2022msmmae, meta2022audiomae, zhong2023ExtendedAudioMAE}. The masked spectrogram modeling principle and architecture design similar to audio MAE \cite{ntt2022msmmae, meta2022audiomae, zhong2023ExtendedAudioMAE} is believed to have contributed to the representation learning potential of SpecMaskGIT. 
Unlike prior arts about finetuning MAE-like architectures for BWE \cite{zhong2023ExtendedAudioMAE, kim2024FrePainter}, SpecMaskGIT enabled BWE in a zero-shot manner.
\vspace{-2mm}
\end{itemize}
We hope this efficient, effective and flexible framework pave the way to the exploration of masked audio modeling toward further diverse scenarios \cite{yang2024visualechoes}. \footnote{Demo:~\url{https://zzaudio.github.io/SpecMaskGIT/index.html}}
\vspace{-0.3cm}
\section{Related Works}
\label{sec:related_works}
\vspace{-0.1cm}
Synthesizing audio signals in raw waveform is challenging and computationally demanding \cite{oord2016wavenet}. 
Therefore, the mainstream approach to audio synthesis is to first generate audio in a compressed latent space, and then restore waveforms from latent representations. 
Auto-regressive models such as Jukebox \cite{dhariwal2020jukebox}, AudioGen \cite{kreuk2022audiogen} and MusicGen \cite{copet2024musicgen} use vector-quantized (VQ) variational auto-encoders (VAE) \cite{oord2017vqvae} to tokenize raw waveforms into a discrete latent space. 
While AudioGen and MusicGen use a higher compression rate than Jukebox, 500 iterations are required to synthesize a 10-second clip, slowing down the speed.

Advances in audio representation learning such as audio MAE (\cite{ntt2022msmmae, meta2022audiomae, zhong2023ExtendedAudioMAE}) indicate that Mel-spectrogram is an effective compression of raw audio signals, as it emphasizes acoustic features of sound events while maintaining sufficient details to reconstruct raw waveforms.
Inspired by the above success of representation learning, several methods used discrete \cite{yang2023diffsound} or continuous \cite{huang2023makeanaudio, liu2023audioldm, ghosal2023tango, huang2023makeanaudio2, liu2023audioldm2} diffusion models upon the latent Mel-spectrogram space created by a VAE or SpecVQGAN \cite{iashin2021SpecVQGAN}. 
These diffusion models require up to 200 iterations for high-fidelity synthesis, which is still challenging for low-resource platforms and interactive use cases.
While distilling a diffusion model can effectively reduce the required iterations \cite{saito2024soundctm}, we limit our discussion to non-distilled methods for a fair comparison.
For Mel-based synthesis methods, waveforms are reconstructed from Mel-spectrogram with a neural vocoder, such as HiFiGAN \cite{kong2020hifigan} or BigVSAN \cite{shibuya2023bigvsan}. 

In pursuit of higher synthesis efficiency, VampNet \cite{garcia2023vampnet} and the concurrent MAGNeT \cite{ziv2024magnet} introduced the parallel iterative synthesis strategy from MaskGIT\cite{chang2022maskgit}. 
MaskGIT, originally proposed for class-conditional image synthesis tasks in \cite{chang2022maskgit}, uses a bi-directional Transformer, instead of the uni-directional counterpart in auto-regressive methods, to reduce the required number of iterations.
Although VampNet and MAGNeT reduced the number of iterations compared to their auto-regressive counterparts, VampNet does not support text prompts, while MAGNeT takes 180 iterations, which is even heavier than some diffusion models that only require 100 iterations \cite{liu2023audioldm, ghosal2023tango, liu2023audioldm2, huang2023makeanaudio2}. 
Moreover, it is difficult for methods built upon wave-domain latent space to address frequency domain tasks such as BWE, limiting their applications.
\vspace{-3mm}
\section{SpecMaskGIT}
\label{sec:proposed_method}
\vspace{-1mm}
The efficiency, effectiveness and flexbility of SpecMaskGIT is the consequence of a combination of efforts, including the high compression rate in the tokenizer, the small model size, fast synthesis algorithm, among others.
\vspace{-2.0mm}
\subsection{Spectrogram Tokenizer and Vocoder}
\label{ssec:specvqgan}
\begin{figure}[tb]
\centerline{\includegraphics[width=0.99\linewidth]{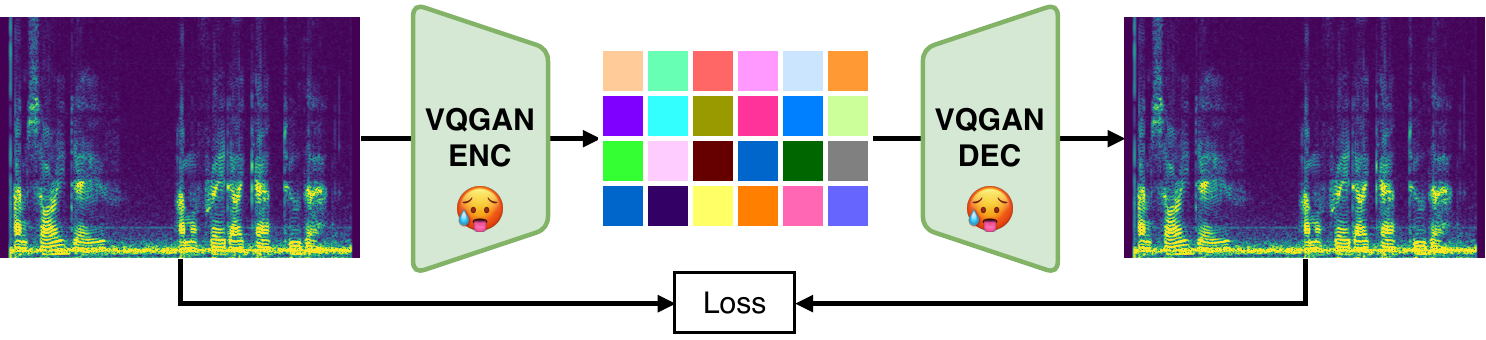}}
\vspace{-0.4cm}
\caption{SpecVQGAN, which encodes non-overlapping 16-by-16 time-mel patches into discrete tokens, and decodes the discrete tokens back to Mel-spectrogram.}
\label{fig:blockdiag-specvqgan}
\vspace{-0.5cm}
\end{figure}

\vspace{-1.0mm}
A modified SpecVQGAN \cite{iashin2021SpecVQGAN} is trained to tokenize non-overlapping 16-by-16 time-mel patches into discrete tokens, and recover the tokens back to Mel-spectrogram as in Fig. \ref{fig:blockdiag-specvqgan}. Reconstructed Mel-spectrograms are then transformed to waveforms by a pre-trained vocoder. On top of the 3.2$\times$ compression offered by the wave-to-mel transform in our configuration, SpecVQGAN further offers 256$\times$ compression of the spectrogram, resulting in total over 800$\times$ compression to the raw waveform, effectively reducing the number of tokens to synthesize.

We utilize the standard Mel transform widely used in vocoders \cite{kumar2019melgan, kong2020hifigan, lee2022bigvgan, shibuya2023bigvsan} for optimal Mel computation, as hyper-parameters of Mel transform has an impact on tokenizer's performance  \cite{huang2023makeanaudio2}. 
To stabilize the training, we keep the spectrogram normalization in the original SpecVQGAN, 
which clips Mel bins lower than -80 dB or louder than 20 dB, and then maps the spectrogram into the range between -1.0 to 1.0. 
Our modified SpecVQGAN is shown competitive in reconstruction quality in Sec. \ref{ssec:res_tta}.
\vspace{-2mm}
\subsection{Masked Generative Modeling of Spectrograms}
\label{ssec:masked_generative_spectrogram_modeling}
\begin{figure}[tb]
\centerline{\includegraphics[width=0.99\linewidth]{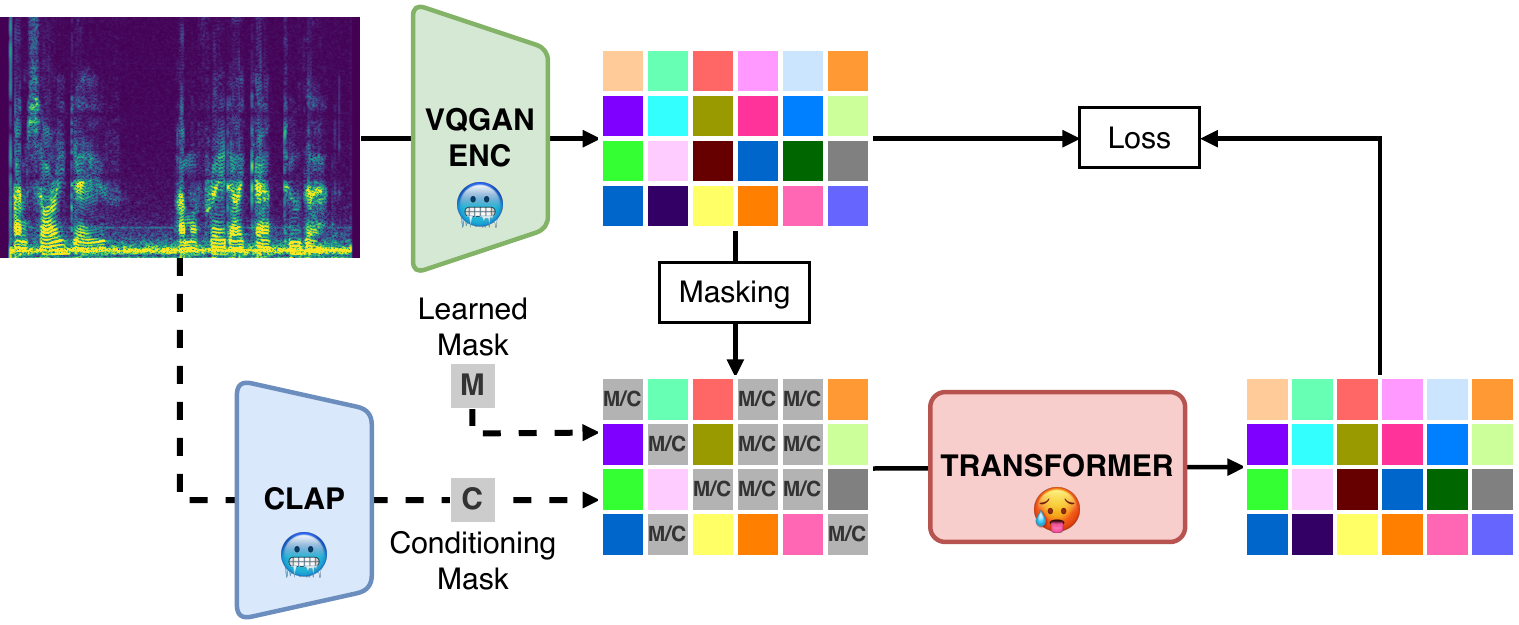}}
\vspace{-0.4cm}
\caption{Self-supervised training of SpecMaskGIT. The Transformer is trained to reconstruct SpecVQGAN token sequences that are randomly masked with variable masking ratios, conditioned by a semantic embeddding from the CLAP encoder. ``M'' denotes the learned mask token, while ``C'' denotes the proposed conditional mask.}
\label{fig:blockdiag-train}
\vspace{-0.4cm}
\end{figure}

\vspace{-1mm}
We train a masked generative Transformer upon the discrete latent space created by the pretrained SpecVQGAN as in Fig. \ref{fig:blockdiag-train}. 
First, the pretrained CLAP encoder maps the input audio to a semantic embedding aligned with its corresponding text descriptions. 
Meanwhile, the input audio is tokenized by SpecVQGAN. 
Finally, similar to representation learning such as audio MAE \cite{ntt2022msmmae, meta2022audiomae, zhong2023ExtendedAudioMAE}, a bi-directional Transformer is trained to reconstruct Mel-spectrogram token sequences from a randomly masked input. 

There are two major differences from audio MAE. 
First, the masking ratio is NOT a fixed value but sampled on-the-fly from a truncated Gaussian distribution that is centered at 55\% \cite{li2023mage} and ranges from 0\% to 100\% \cite{chang2022maskgit}. 
As a result, although in each training step SpecMaskGIT behaves similarly to audio MAE, it learns the training data distribution from various masking ratios, hence gaining the ability to iteratively refine audio tokens by gradually decreasing the masking ratio across multiple iterations, which is explained in Sec.~\ref{ssec:paralle_iterative_decoding}. 
The other difference lies in the loss function. 
Audio MAE works on raw Mel-spectrograms, thus the mask reconstruction is optimized by mean square error. 
However, SpecMaskGIT works in a discrete latent space, which means the reconstruction of a masked position evolves to the retrieval of a correct code from the codebook of SpecVQGAN, 
\textit{i.e.}, a multi-class single-label classification procedure. 
Therefore, the loss function becomes the cross entropy (CE) loss with label smoothing equal to 0.1.
Following audio MAE, those visible positions in the input are not considered in the loss calculation:
\vspace{-1.5mm}
\begin{equation}
    \textrm{Loss} = \textrm{CE}(\textrm{prediction}[\textrm{mask}], \textrm{label}[\textrm{mask}]).
\vspace{-1mm}
\end{equation}
\begin{figure*}[tb]
\centerline{\includegraphics[width=0.85\linewidth]{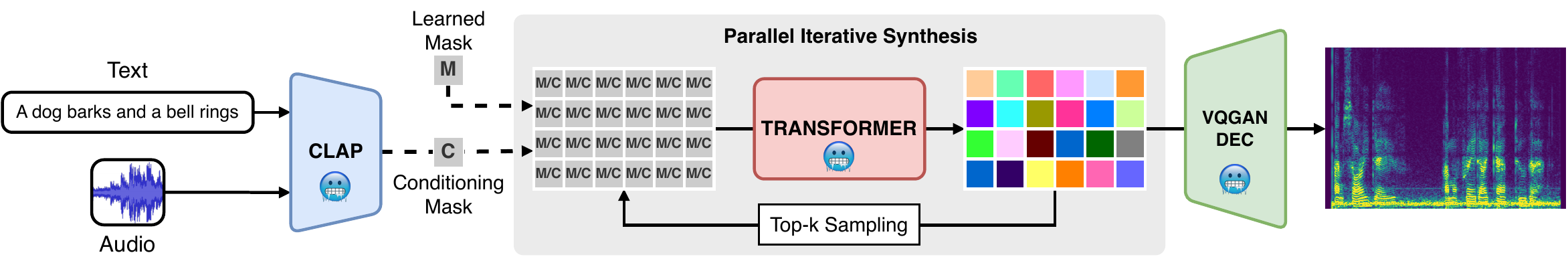}}
\vspace{-6mm}
\caption{The iterative text-to-audio synthesis in SpecMaskGIT. }
\label{fig:blockdiag-inference}
\vspace{-0.3cm}
\end{figure*}
\begin{figure*}[tb]
\centerline{\includegraphics[width=0.80\linewidth]{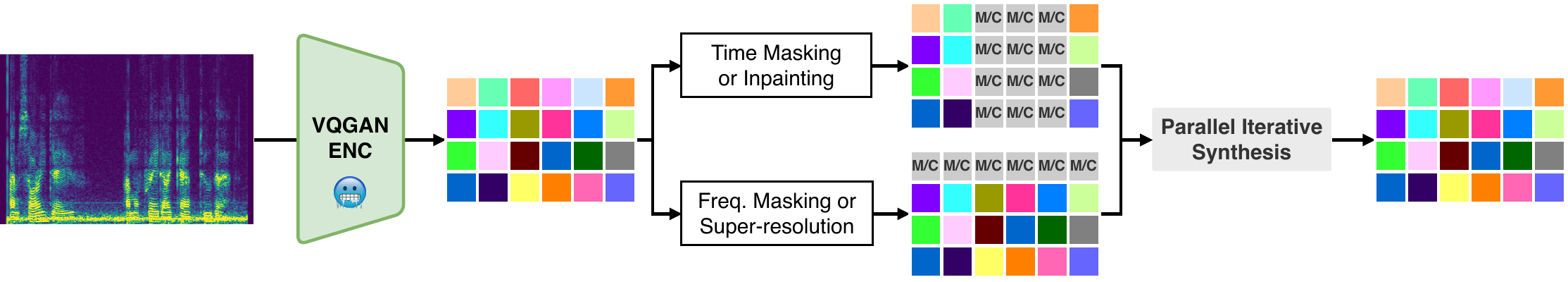}}
\vspace{-0.5cm}
\caption{Zero-shot time inpainting and bandwidth extension for general audio data via SpecMaskGIT.}
\label{fig:blockdiag-inpaint}
\vspace{-0.5cm}
\end{figure*}
\vspace{-7mm}
\subsection{Text Conditioning via Sequential Modeling}
\label{ssec:text_condition}
\vspace{-1mm}
Similarly to \cite{liu2023audioldm}, we train SpecMaskGIT without audio-text pairs by using a pretrained CLAP model \cite{laion2023clap}, for which audio and text embeddings are aligned in a shared latent space.
Leveraging such alignment, after training with the audio branch of CLAP in Fig.~\ref{fig:blockdiag-train}, we can directly condition our pretrained model with the text branch as in Fig.~\ref{fig:blockdiag-inference}.
We use a publicly available CLAP checkpoint (``630k-audioset-best.pt''  \cite{laion2023clap}) for better reproducibility.

Although the above design is inspired by AudioLDM \cite{liu2023audioldm}, SpecMaskGIT is different in the way to inject CLAP conditions. 
Besides the FiLM mechanism (\cite{perez2018film}) used in AudioLDM, 
prior arts inject text conditions into the generative model via the cross-attention mechanism \cite{yang2023diffsound, huang2023makeanaudio, ghosal2023tango, huang2023makeanaudio2, liu2023audioldm2}, 
even for methods based on sequential modeling such as AudioGen \cite{kreuk2022audiogen} and MAGNeT \cite{ziv2024magnet}, 
which inevitably involves efforts to modify basic DNN modules.
We believe that reusing identical DNN modules, such as the Vision Transformer (ViT) \cite{dosovitskiy2020vit}, across different tasks is beneficial to efficient development, 
so we choose to achieve text-conditional audio synthesis by pure sequential modeling, 
\textit{i.e.}, appending the CLAP embedding to the input sequence of the Transformer.
As a result, SpecMaskGIT can be implemented by the same ViT used in audio MAE \cite{ntt2022msmmae, meta2022audiomae, zhong2023ExtendedAudioMAE}, and thus we interpret SpecMaskGIT as a generative extension to previous discriminative ways of masked spectrogram modeling. We hypothesize the masked spectrogram modeling and ViT implementation similar to audio MAE has contributed to the representation learning potential of SpecMaskGIT, as is shown in Sec.~\ref{ssec:res_downstream}, 

While the common practice in \cite{ntt2022msmmae, meta2022audiomae, zhong2023ExtendedAudioMAE, chang2022maskgit} is to use a learnable but input-independent token to indicate which parts in the sequence are masked (``M'' in Fig. \ref{fig:blockdiag-train}), 
the mask reconstruction task is challenging as the input-independent mask offers no hint for a better reconstruction.
To further guide the mask reconstruction procedure, we propose to directly use the input-dependent CLAP embedding as a conditional mask (``C'' in Fig. \ref{fig:blockdiag-train}), which offers semantic hints like ``a dog barking sound'' to the model, and is found beneficial to TTA performance in Sec.~\ref{ssec:res_tta}. 
\vspace{-1mm}
\subsection{Iterative Synthesis with Classifier-free Guidance}
\label{ssec:paralle_iterative_decoding}
We follow the parallel iterative synthesis strategy proposed in MaskGIT \cite{chang2022maskgit} in general, 
but employ classifier-free guidance (CFG) \cite{ho2022classifier} to improve the synthesis quality.
This iterative algorithm allows SpecMaskGIT to synthesize multiple high-quality tokens at each iteration, reducing the number of iterations to a value one order-of-magnitude smaller than previous TTA methods.

To enable CFG, we replace CLAP embedding with the learned mask token on a random 10\% of training steps. 
At inference phase, both the conditional logit $\ell_{c}$ and unconditional logit $\ell_{u}$ for each masked token are computed. 
The final logits $\ell_{g}$ are made by a linear combination of the two logits based on $t$, the guidance scale:
\begin{equation}
\label{eq:cfg}
    \ell_{g} = \ell_{u} +  t(\ell_{c} - \ell_{u}).
\end{equation} 
Intuitively, CFG balances between diversity and audio-text alignment. 
Inspired by \cite{chang2023muse}, we introduce a linear scheduler to the guidance scale $t$, which linearly increases $t$ from 0.0 to an assigned value through the synthesis iterations. This allows the result of early iterations to be  more diverse (unconditional) with low guidance, but increases the influence of the conditioning for the later synthesis, and is proved beneficial to synthesis quality in Sec. \ref{ssec:res_tta}.

The parallel iterative synthesis of SpecMaskGIT shown in Fig. \ref{fig:blockdiag-inference} is explained as follows.

\noindent\textbf{1. Estimating.} The Transformer estimates the probability of being the correct code at each masked position for all discrete codes in the SpecVQGAN codebook.

\noindent\textbf{2. Unmasking.} Given the probabilities over the codebook for each masked position, a code is retrieved based on the categorical sampling to unmask that position. This step is different from the deterministic unmasking in audio MAE. 

\noindent\textbf{3. Scheduling.} Although SpecMaskGIT can unmask all positions at once, the quality of the synthesized audio is low. 
To iteratively refine the synthesis, we need to re-mask the result to a masking ratio that is lower than the current iteration. 
We follow the common practice in \cite{chang2022maskgit, li2023mage, garcia2023vampnet, ziv2024magnet} to use a cosine scheduler to decide the masking ratio in each iteration. 
The cosine scheduler re-masks a larger portion of the synthesized audio for the early iterations, which is intuitive as the quality in earlier iterations is lower.

\noindent\textbf{4. Top-$k$ sampling}. Given the masking ratio for the next iteration, we get to know $k$ tokens are going to be re-masked. 
The log-likelihood of unmasked tokens is used to decide the $k$ worst tokens. Since it is observed that a deterministic top-k retrieval leads to the synthesis of monotonic images in \cite{besnier2023maskgit_reproduce}, we followed \cite{li2023mage, garcia2023vampnet} to add a Gumbel noise to log-likelihood, making the top-$k$ sampling stochastic: 
\vspace{-1.5mm}
\begin{equation}
    \textrm{confidence} = \log(p) + t_\textrm{gumbel} \cdot n_\textrm{gumbel},
\end{equation}
where $p$ is the probabilities of all unmasked tokens calculated from the CFG logits in Eq. \ref{eq:cfg}, $n_\textrm{gumbel}$ is Gumbel noise, 
and $t_\textrm{gumbel}$ is the temperature multiplied to Gumbel noise. Following \cite{li2023mage}, we linearly anneal the $t_\textrm{gumbel}$ by a coefficient defined as $\textrm{iter}/\textrm{num\_iter}$, 
where ``iter'' means the index of the current iteration, ``num\_iter'' stands for the number of all scheduled iterations.

\noindent\textbf{5. Repeating.} Repeat the above operations until the cosine scheduler reduces the masking ratio to 0.

For TTA, the SpecMaskGIT starts the above iterative procedure from a fully masked sequence as in Fig.~\ref{fig:blockdiag-inference}. 
Meanwhile, the iterative algorithm is also valid when the masking ratio of input sequence is lower than 100\%, 
which automatically enables zero-shot inpainting in both time and frequency domain as is shown in Fig.~\ref{fig:blockdiag-inpaint}. 
It is worth noticing that since VampNet \cite{garcia2023vampnet} and MAGNeT \cite{ziv2024magnet} employ a wave-domain tokenizer, explicit frequency inpainting (BWE) is difficult.
\vspace{-1mm}
\section{Experiments}
\label{sec:exp}
We pretrained the SpecVQGAN \cite{iashin2021SpecVQGAN} and two vocoders (HiFiGAN \cite{kong2020hifigan} \& BigVSAN \cite{lee2022bigvgan}) on AudioSet (AS) unbalanced and balanced subset \cite{gemmeke2017audioset} for 1.5M steps. 
The AS we collected contains around 1.8 million 10-second audio segments of diverse sound sources and recording environments. AS has been widely used in general audio representation learning \cite{ntt2022msmmae, meta2022audiomae, zhong2023ExtendedAudioMAE}. 
We followed the ``VGGSound'' configuration in the original SpecVQGAN repository \cite{iashin2021SpecVQGAN} without using LPAPS loss as suggested in the repository. 
Our SpecVQGAN has around 75M parameters, and a codebook of 1024 codes, each of which is represented by a 256-dim embedding. 
As mentioned in Sec. \ref{ssec:specvqgan}, the standard Mel-spectrogram transform from vocoders \cite{kong2020hifigan, shibuya2023bigvsan} is utilized, which transforms a 10-second audio clip at sampling rate 22.05kHz into 848 frames with 80 Mel bins. The Mel-spectrogram is further tokenized into 265 tokens.

SpecMaskGIT employs the ViT implementation widely used in previous audio masked Transformers \cite{rw2019timm, koutini2021patchout, ntt2022msmmae, zhong2023ExtendedAudioMAE}. 
To be consistent with the image MaskGIT \cite{chang2022maskgit}, 24 Transformer blocks are used, in which the attention dimension is 768 with 8 heads and the feedforward dimension is 3072, resulting in around 170M parameters.
We trained SpecMaskGIT on AS for 500k steps with a batch size of 112. When training the model on AudioCaps (AC) \cite{kim2019audiocaps}, 
we train for 250k steps with a batch size of 48, as AC only contains 50k 10-second audio clips.  
To stably train SpecMaskGIT, we follow the common practice in \cite{ntt2022msmmae, meta2022audiomae, zhong2023ExtendedAudioMAE} to employ a linear warmup and then a cosine annealing of the learning rate (LR). We warmup 16k steps for AS and 5k steps for AC. The base LR is set to 1e-3, and the LR equates to the division of base LR by batch size \cite{meta2022audiomae, li2023mage}.
The iterative synthesis algorithm is based on the open-source implementation of \cite{li2023mage}.

To evaluate the \textbf{text-to-audio synthesis} quality of SpecMaskGIT, we benchmark on the AudioCaps (AC) test set with the text prompts released by \cite{liu2023audioldm} for fair comparison.
To investigate the flexibility of SpecMaskGIT in downstream tasks, we use the SpecMaskGIT trained on AS for 500k steps in the following tasks:
\textbf{Zero-shot time inpainting}. we manually mask out the 25th to 35th Mel-spec frames (around 1.9s) of AC test set, 
and employ the SpecMaskGIT to inpaint the lost regions in a zero-shot manner, 
\textit{i.e.}, inpainting without any task-specific finetuning. 
\textbf{Zero-shot audio bandwidth extension}: The top 16 Mel-spec bins (\textit{i.e.}, components beyond 4.3kHz) of AC test set are masked, which creates a 2.5$\times$ BWE task.
For all tasks above, we use the toolbox in \cite{comunita2019fad_python} to compute FAD (\cite{kilgour2018fad}) scores as the metric, 
since FAD has been widely used to evaluate TTA \cite{liu2023audioldm, ghosal2023tango, huang2023makeanaudio2, liu2023audioldm2}, time inpainting \cite{liu2023audioldm} and BWE \cite{moliner2024BWE} tasks. 
To investigate the representation learning potential of SpecMaskGIT, we further linear probe the model for the \textbf{music tagging} task in MagnaTagATune (MTAT) dataset \cite{LawWMBD09mtat} with ROC-AUC and mAP as metrics \cite{li2023mert}. MTAT presents a multi-label task for genre, instrument and mood, thus has been widely used to evaluate music tagging models \cite{li2023mert, janne2021clmr, pons2019musicnn, McCallumKOGE22mule}. 
We use a single linear layer with batch normalization and 0.1 dropout as the probe.
\vspace{-3.0mm}
\section{Results}
\label{sec:results}
\vspace{-0.5mm}
\subsection{Text-to-audio Synthesis}
\label{ssec:res_tta}
\vspace{-1mm}
We report the FAD scores of SpecMaskGIT in Tab.~\ref{tab:benchmark_ac_discrete} together with other discrete models. 
Our model is first trained on AS for 500k steps and then finetuned on AC train set for 250k steps. The CFG scale is set to 3.0 empirically.
SpecMaskGIT outperforms Diffsound (VQ-Diffusion), MAGNeT-small (similar to SpecMaskGIT but in latent wave domain), as well as AudioGen-base (auto-regressive) in terms of FAD with one order-of-magnitude fewer iterations. 
The FAD score is achieved without training with any audio-text pairs, which proved the effectiveness of such self-supervised training for discrete models. We also found the proposed conditional mask explained in Sec.~\ref{ssec:text_condition} improves the FAD score without additional parameter or computation. Both the CFG and linear scheduler of CFG scale contributed to the FAD.

Given the small number of iterations and small model size, SpecMaskGIT can synthesize realistic 10-second audio clips in real-time with only 4 cores of a Xeon CPU as is shown in Fig.~\ref{fig:runtime_cpu}, or even 30$\times$ faster than real-time on one RTX-A6000 GPU. The efficiency and effectiveness of SpecMaskGIT make the model attractive to interactive applications and low-resource environments.
\begin{table}[t] 
\centering
\caption{Comparing SpecMaskGIT with other discrete TTA methods on AudioCaps test set.}
\label{tab:benchmark_ac_discrete}
\resizebox{0.9\linewidth}{!}{
\begin{tabular}{lcccc}
\toprule
    Method & Params & Text & Num\_iter & FAD\\
\midrule
    Diffsound \cite{yang2023diffsound} & 400M & Yes & 100 & 7.8 \\
    MAGNeT-small \cite{ziv2024magnet} & 300M & Yes & 180 & 3.2 \\
    AudioGen-base \cite{kreuk2022audiogen} & 285M & Yes & 500 & 3.1\\
    AudioGen-large \cite{kreuk2022audiogen} & 1.5B & Yes & 500 & \textbf{1.8}\\
\hline
    SpecMaskGIT (ours) & \multirow{5}*{\textbf{170M}} & \multirow{5}*{\textbf{No}} & \multirow{5}*{\textbf{16}} & 2.7\\
    - w HiFiGAN &&&& 2.8 \\
    \ \ - w/o conditional mask &&&& 3.2 \\
    \ \ - w/o CFG &&&& 3.1 \\
    \ \ - w/o CFG linear scheduler &&&& 3.1 \\
\bottomrule
    \end{tabular}}
\label{tab:rFAD}
\vspace{-3.5mm}
\end{table}

When compared to state-of-the-art (SOTA) continuous diffusion models in Tab.~\ref{tab:res_ac_all}, SpecMaskGIT could not achieve a comparable FAD score, 
but we emphasize that the proposed method offers decent performance with high efficiency, 
\textit{i.e.}, smaller model size and fewer iterations, which can be clearly seen in Fig.~\ref{fig:fad_numiter_size}. 
Overall, continuous methods are more advantageous in FAD than discrete methods. We leave the further improvement of discrete generative model as future work.

\begin{figure}[h]
\vspace{-3mm}
    \centering
    \begin{subfigure}
        {\includegraphics[height=0.90in]{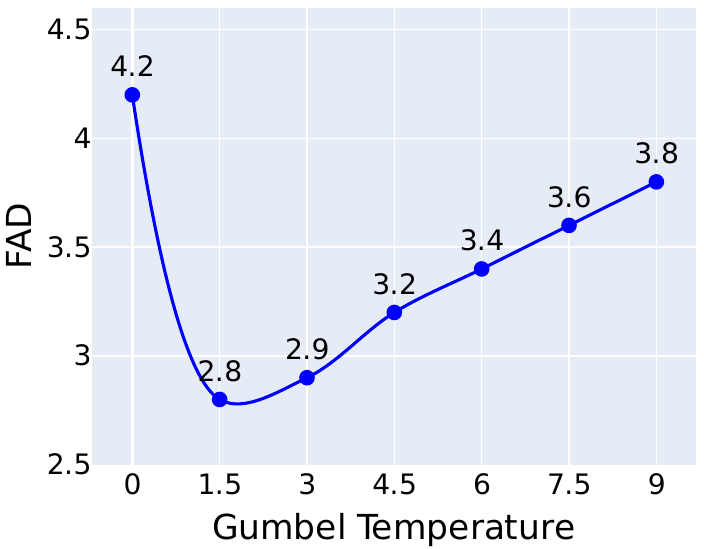}}
    \end{subfigure}
    \begin{subfigure}
        {\includegraphics[height=0.90in]{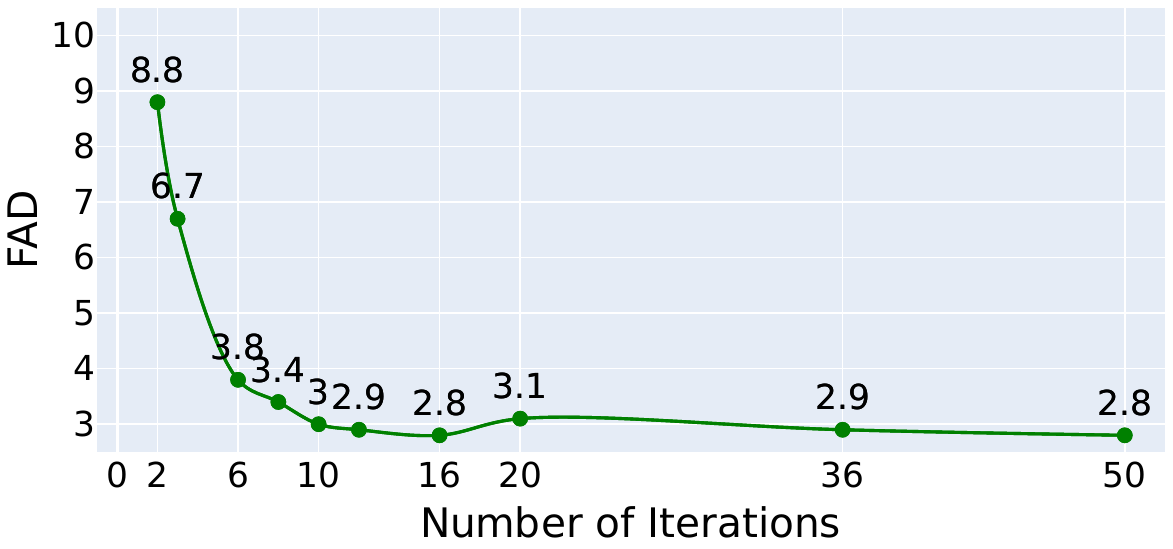}}
    \end{subfigure}
    \vspace{-4mm}
    \caption{Left: FAD vs. Gumbel temperature. Right: FAD vs. Number of iterations.}
    \label{fig:fad_gumbel_temp_num_iter}
    \vspace{-1mm}
\end{figure}

\begin{table}[tb] 
\centering
\caption{Benchmarking on AudioCaps test set. Dis.: discrete methods. Con.: continuous methods.}
\label{tab:res_ac_all}
\resizebox{7.5cm}{!}{
\begin{tabular}{c|c|cc|c|c}
\toprule
    Method & Params &Dis.&Con.& Num\_iter & FAD\\
\midrule
    Diffsound \cite{yang2023diffsound} & 400M & \Checkmark & & 100 & 7.8 \\
    Make-an-Audio \cite{huang2023makeanaudio} & 330M & & \Checkmark & 100 & 4.6\\
    MAGNeT-small \cite{ziv2024magnet} & 300M & \Checkmark & & 180 & 3.2 \\
    AudioGen-base \cite{kreuk2022audiogen} & 285M & \Checkmark & & 500 & 3.1\\
    AudioLDM-Medium-full-FT \cite{liu2023audioldm} & 420M & & \Checkmark & 100 & 2.6\\
    AudioLDM-Large-full-FT \cite{liu2023audioldm} & 740M & & \Checkmark & 200 & 2.0\\
    Make-an-Audio 2 \cite{huang2023makeanaudio2} & 940M & & \Checkmark & 100 & 1.8\\
    AudioGen-large \cite{kreuk2022audiogen} & 1.5B & \Checkmark & & 500 & 1.8\\
    AudioLDM2-Small-AC \cite{liu2023audioldm2} & 350M & & \Checkmark & 200 & 1.7\\
    TANGO-AC \cite{ghosal2023tango} & 870M & & \Checkmark & 100 & 1.6\\
    AudioLDM2-Large-AC \cite{liu2023audioldm2} & 710M & & \Checkmark & 200 & \textbf{1.4}\\
\hline
    SpecMaskGIT (ours) & \textbf{170M} & \Checkmark & & \textbf{16} & 2.7\\
\bottomrule
    \end{tabular}}
\label{tab:rFAD}
\vspace{-3mm}
\end{table}
\noindent\textbf{Ablation study: Gumbel noise and number of iterations in SpecMaskGIT.} We use HiFiGAN in all ablation studies. As mentioned in Sec.~\ref{ssec:paralle_iterative_decoding}, the Gumbel noise is essential to the top-$k$ sampling during the iterative synthesis. Fig. \ref{fig:fad_gumbel_temp_num_iter} shows that a temperature of 1.5 is the optimal. 
SpecMaskGIT achieves decent performance (FAD = 3.4) with only 8 iterations, and reaches its best (FAD = 2.8) with 16 iterations. More iterations do not improve the performance, which is consistent with the image MaskGIT \cite{chang2022maskgit}.

\noindent\textbf{Ablation study: Audio reconstruction quality.} We evaluate the reconstruction FAD (rFAD) scores of two vocoders and the SpecVQGAN in Tab.~\ref{tab:rFAD} with previous methods reported in \cite{huang2023makeanaudio2}. 
Even with a similar architecture, rFAD of DiffSound and SpecMaskGIT can vary a lot due to different Mel computation and vocoder. 
Our pipeline achieves SOTA level rFAD scores for Mel-spectrogram methods while maintaining the highest compression rate or the lowest latent rate, which helped SpecMaskGIT to outperform methods such as Diffsound and Make-an-audio by a large margin yet with higher efficiency. 
We further analyze the rFAD of vocoders by inputting ground truth Mel to them, and found a significant performance gap between HiFiGAN and BigVSAN, which is not observed when vocoders are combined with SpecVQGAN. This indicates that SpecVQGAN has been the bottleneck in reconstruction quality and asks for future improvements.
\begin{table}[h] 
\centering
\vspace{-5mm}
\caption{rFAD of Mel-spectrogram VAEs and Vocoders on AudioCaps test set. \textbf{Bold}: best overall rFAD.}
\label{tab:rFAD}
\resizebox{1.0\linewidth}{!}{
\begin{tabular}{ccccc}
\toprule
    Method & Mel-spec VAE & Vocoder & Latent rate & rFAD\\
\midrule
    Diffsound \cite{yang2023diffsound}  & SpecVQGAN & MelGAN & 27Hz & 6.2 \\
    Make-an-audio \cite{huang2023makeanaudio}  & VAE-GAN & HiFiGAN & 78Hz & 6.0 \\
    AudioLDM \cite{liu2023audioldm} & VAE-GAN & HiFiGAN & 410Hz & 1.2\\
    Make-an-audio 2 \cite{huang2023makeanaudio2} & VAE-GAN & BigVGAN & 31Hz & \textbf{1.0}\\
\hline
    \multirow{4}*{SpecMaskGIT (ours)} & - & \multirow{2}*{HiFiGAN} & \multirow{2}*{27Hz} & 0.4\\
     & SpecVQGAN & & & 1.1\\
\cline{2-5}
     & - & \multirow{2}*{BigVSAN} & \multirow{2}*{27Hz} & 0.1\\
     & SpecVQGAN & & & \textbf{1.0}\\
\bottomrule
    \end{tabular}}
\label{tab:rFAD}
\vspace{-1mm}
\end{table}

\noindent\textbf{Ablation study: Bias in AudioCaps benchmark.} 
The dataset gap between AC and other larger \& more diverse datasets is investigated. 
It is observed in \cite{liu2023audioldm} that finetuning (FT) a TTA model on AC improves the TTA performance in terms of FAD, though the model is pretrained on a larger dataset. 
We reproduced this phenomenon with SpecMaskGIT as shown in Tab. \ref{tab:influence_finetune}. 
We also observed that training on the small-scale AC alone brought better FAD score than the model trained with larger datasets in Tab. \ref{tab:influence_audiocaps_from_scratch}, 
which is consistent with \cite{ghosal2023tango, liu2023audioldm2}.
\begin{table}[t] 
\centering
\caption{AudioCaps test set performance before and after AudioCaps finetuning (FT).}
\label{tab:influence_finetune}
\resizebox{7.5cm}{!}{
\begin{tabular}{ccccc}
\toprule
    \multirow{2}*{Method} & \multirow{2}*{Params} & \multirow{2}*{Num\_iter} & \multicolumn{2}{c}{FAD}\\
    & & & before FT & after FT\\
\midrule
    AudioLDM-Small-full \cite{liu2023audioldm} & 180M & 200 & 4.9 & 2.3\\
    AudioLDM-Large-full \cite{liu2023audioldm} & 740M & 200 & 4.2 & 2.0\\
\hline
    SpecMaskGIT (ours) & 170M & 16 & 4.2 & 2.8\\
\bottomrule
    \end{tabular}}
\label{tab:rFAD}    
\end{table}
\begin{table}[t] 
\centering
\vspace{-5mm}
\caption{Small-scale AudioCaps training results in better scores than large-scale dataset.}
\label{tab:influence_audiocaps_from_scratch}
\resizebox{0.92\linewidth}{!}{
\begin{tabular}{ccccc}
\toprule
    \multirow{2}*{Method} & \multirow{2}*{Params} & \multirow{2}*{Num\_iter} & \multicolumn{2}{c}{FAD}\\
    & & & Other datasets & AudioCaps\\
\midrule
    AudioLDM-Small \cite{liu2023audioldm} & 180M & 200 & 4.9 & 2.4\\
    AudioLDM-Large \cite{liu2023audioldm} & 740M & 200 & 4.2 & 2.1\\
    AudioLDM2-Small \cite{liu2023audioldm2} & 350M & 200 & 2.1 & 1.7\\
    AudioLDM2-Large \cite{liu2023audioldm2} & 710M & 200 & 1.9 & 1.4\\
\hline
    SpecMaskGIT (ours) & 170M & 16 & 4.2 & 2.9\\
\bottomrule
    \end{tabular}}
\label{tab:rFAD}
\vspace{-3mm}
\end{table}

\begin{figure}[htb]
\vspace{-1mm}
\centerline{\includegraphics[width=0.9\linewidth]{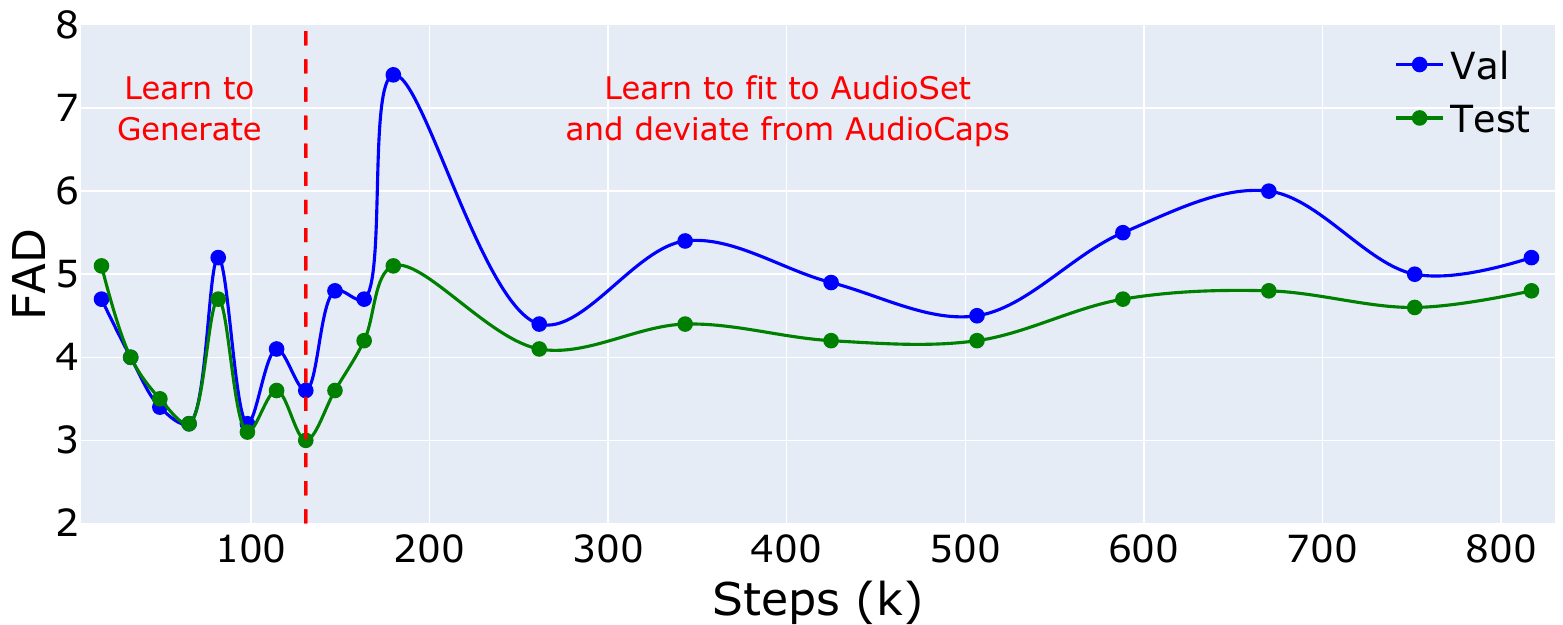}}
\vspace{-0.4cm}
\caption{FAD vs. AudioSet training steps. }
\label{fig:fad_vs_training_step}
\vspace{-0.3cm}
\end{figure}

We hypothesize that there is a data distribution gap between AC and other datasets, 
such that when a model fully fits other datasets, the distribution of its synthesis deviates from AC, resulting in worse FAD. 
Therefore, we continued to train SpecMaskGIT on AS until 800k steps, 
and depict the ``FAD vs. training step'' curves on both the valid and test set of AC to verify our hypothesis. 
It is clear in Fig. \ref{fig:fad_vs_training_step} that SpecMaskGIT learns to synthesize audio in the early stage and keeps improving the FAD on AC. 
As the training goes on, SpecMaskGIT just fits toward AS, which worsens the FAD in AC. 

Inspired by audio classification tasks in which early stop is applied to prevent the model from overfitting to the train set distribution, 
we proposed to apply early stop to the SpecMaskGIT model trained solely on AS, 
and report the competitive FAD score with other methods that are without AC finetuning of AC-alone training in Tab. \ref{tab:no_finetune}. 
We believe that a more comprehensive and less biased benchmark will contribute to the future advances of TTA research.
\begin{table}[tb] 
\centering
\caption{Benchmarking on AudioCaps test set without AC finetuning or AC-alone training.}
\resizebox{0.9\linewidth}{!}{
\begin{tabular}{c|c|cc|c|c}
\toprule
    Method & Params &Dis.&Con.& Num\_iter & FAD\\
\midrule
    Diffsound \cite{yang2023diffsound} & 400M & \Checkmark & & 100 & 7.8 \\
    AudioLDM-Small-full \cite{liu2023audioldm} & 180M & & \Checkmark & 200 & 4.9\\
    Make-an-Audio \cite{huang2023makeanaudio} & 330M & & \Checkmark & 100 & 4.6\\
    AudioLDM-Large-full \cite{liu2023audioldm} & 740M & & \Checkmark & 200 & 4.2\\
    MAGNeT-small \cite{ziv2024magnet} & 300M & \Checkmark & & 180 & 3.2 \\
    AudioGen-base \cite{kreuk2022audiogen} & 285M & \Checkmark & & 500 & 3.1\\
    AudioLDM2-Small-full \cite{liu2023audioldm2} & 350M & & \Checkmark & 200 & 2.1\\
    AudioLDM2-Large-full \cite{liu2023audioldm2} & 710M & & \Checkmark & 200 & 1.9\\
    Make-an-Audio 2 \cite{huang2023makeanaudio2} & 940M & & \Checkmark & 100 & \textbf{1.8}\\
    AudioGen-large \cite{kreuk2022audiogen} & 1.5B & \Checkmark & & 500 & \textbf{1.8}\\
\hline
    SpecMaskGIT-AS-EarlyStop (ours) & \textbf{170M} & \Checkmark & & \textbf{16} & 2.9\\
\bottomrule
    \end{tabular}}
\label{tab:no_finetune}    
\vspace{-5mm}
\end{table}
\vspace{-2mm}
\subsection{Downstream Inpainting, BWE and Tagging Tasks}
\label{ssec:res_downstream}
\vspace{-1mm}
Results of the time inpainting and audio BWE tasks are shown in Tab. \ref{tab:BWE}. 
We utilized the pipeline in Fig.  \ref{fig:blockdiag-inpaint} unconditionally, with Gumbel temperature 1.5 and 16 iterations. 
SpecMaskGIT significantly improved the input signals in terms of FAD, validating its zero-shot ability in such tasks. 
BWE performance can be further improved by applying the low-frequency replacement (LFR) technique \cite{liu2022nvsr, liu2024audiosr}. 
Unlike prior arts that finetune MAE-like architectures for BWE \cite{zhong2023ExtendedAudioMAE, kim2024FrePainter}, SpecMaskGIT achieves it by zero-shot.
\begin{table}[t] 
\centering
\caption{Zero-shot time inpainting and BWE FAD scores.}
\resizebox{0.60\linewidth}{!}{
\begin{tabular}{lcc}
\toprule
 & BWE & Time inpaint\\
\midrule
    Unprocessed & 2.7 & 1.6\\
\hline
    SpecMaskGIT (ours) & 1.5 & \textbf{1.2}\\
    - w/ LFR & \textbf{0.4} & - \\
\hline
    Ground truth & 0.0 & 0.0\\
\bottomrule
    \end{tabular}}
\label{tab:BWE}
\vspace{-3mm}
\end{table}

In Tab.~\ref{tab:tagging}, the potential of SpecMaskGIT in representation learning is confirmed by the music tagging performance on MTAT dataset. As a TTA model, SpecMaskGIT outperforms classification-specialized models such as CLMR, MusiCNN, MULE, and MERT (the MAE-like model in wave domain). SpecMaskGIT got an ROC-AUC comparable to Jukebox which contains 5B parameters. We hypothesize the tagging capability comes from the masked spectrogram modeling and ViT implementation  similar to audio MAE, as explained in Sec. \ref{sec:proposed_method}.
\begin{table}[h] 
\centering
\vspace{-5mm}
\caption{Music tagging performance on MTAT.}
\resizebox{\linewidth}{!}{
\begin{tabular}{c|ccccc|c}
\toprule
    \multirow{2}*{Method} & CLMR & MusiCNN & MERT- & MULE- & Jukebox & SpecMask-\\
     & \cite{janne2021clmr} & \cite{pons2019musicnn} & 330M \cite{li2023mert} & contrastive \cite{McCallumKOGE22mule} & \cite{dhariwal2020jukebox, castellon2021calm} & GIT\\
\midrule
    mAP (\%) & 36.1 & 38.3 & 40.2 & 40.4 & \textbf{41.4} & 40.5\\
    ROC-AUC (\%) & 89.4 & 90.6 & 91.3 & 91.4 & \textbf{91.5} & \textbf{91.5}\\
\bottomrule
    \end{tabular}}
\label{tab:tagging}
\vspace{-2mm}
\end{table}

We leave the in-depth investigation of SpecMaskGIT in downstream tasks as future work.
\vspace{-2mm}
\section{Conclusion}
Generative models that iteratively synthesize audio clips sparked great success to text-to-audio synthesis (TTA). However, due to hundreds of iterations required in the inference phase and large amount of model parameters,  high-quality TTA systems remain inefficient.
To address the challenges, we propose SpecMaskGIT, a light-weighted, efficient yet effective TTA model based on the masked generative modeling of spectrograms. 
SpecMaskGIT synthesizes realistic audio clips by less than 16 iterations, an order-of-magnitude less than previous iterative TTA methods. It also outperforms larger discrete models in the TTA benchmark, while being real-time with 4 CPU cores or even 30$\times$ faster with a GPU. 
Compared to similar methods, SpecMaskGIT is more flexible in downstream tasks such as zero-shot bandwidth extension.
Moreover, we interprete SpecMaskGIT as a generative extension to audio MAE and shed light on its audio representation learning potential.
We hope our work inspires the exploration of masked audio modeling toward further diverse scenarios.
\clearpage
\bibliography{ISMIRtemplate}
\end{document}